# Multi-model evidence of future tropical Atlantic precipitation change modulated by AMOC decline

Giada Cerato[1*], Katinka Bellomo[1,2], Roberta D'Agostino[3], and Jost von Hardenberg[1,2]

[1] *Politecnico di Torino, Department of Environment, Land, and Infrastructure Engineering, Turin, Italy*

[2] *National Research Council, Institute of Atmospheric Sciences and Climate, Turin, Italy*

[3] *National Research Council, Institute of Atmospheric Sciences and Climate, Lecce, Italy*

* Corresponding author: Giada Cerato, giada.cerato@polito.it

# ABSTRACT

Projections from global climate models reveal a significant inter-model spread in future rainfall changes in the tropical Atlantic by the end of the 21$^{st}$ century, including alterations to the Intertropical Convergence Zone (ITCZ) and monsoonal regions. While existing studies have identified various sources of uncertainty, our research uncovers a prominent role played by the decline of the Atlantic Meridional Overturning Circulation (AMOC) for the inter-model spread. Firstly we examine 30 climate model simulations (using the ssp5-8.5 scenario) from the CMIP6 archive and show that models that present a more substantial AMOC decline exhibit an equatorward shift of the ascending branch of the Atlantic regional Hadley circulation, resulting in a southward displacement of the ITCZ. Conversely, models characterized by a smaller AMOC decline do not indicate any ITCZ displacement. Secondly, we use targeted experiments (using the abrupt 4x$CO_2$ experiment) to specifically isolate the effects of a weakened AMOC from the changes in precipitation that would occur if, under continuous global warming, the AMOC did not weaken. Our results demonstrate that net precipitation anomalies in the abrupt 4x$CO_2$ experiments are displaced southwards compared to the simulation with fixed AMOC strength, corroborating our previous findings. Our study has implications for understanding the mechanisms driving future changes in tropical Atlantic precipitation, and underscores the central role played by the AMOC in future climate change.

## 1. Introduction

The Intertropical Convergence Zone (ITCZ) is a planetary-scale zonal band of intense precipitation activity. Typically, the ITCZ is located about 5° north of the equator in the annual mean (Philander et al. (1996)). However, its location undergoes migration in response to solar insolation and inter-hemispheric energy imbalance (Broccoli et al., (2006); Chiang et al. (2012), Bischoff and Schneider (2014)). Modifications in the strength, width, and seasonal displacements of the ITCZ can have far-reaching impacts, including severe droughts or extensive flooding, with significant implications for local communities (Fernandes et al. (2011), Marengo et al. (2012), Chiang (2002)). Atmospheric reanalyses and observations indicate recent intensification and narrowing of the ITCZ. For the coming years, idealized climate model simulations of future climate change generally project further narrowing and weakening of the mean ascent (Byrne et al. (2018), Mamalakis et al. (2021), Byrne and Schneider (2016)). Nonetheless, previous studies have revealed significant inter-model uncertainty in simulating the response of tropical precipitation to future climate change. For instance, Byrne et al. (2018) demonstrated inter-model uncertainty in the meridional shift of the ITCZ in a warming climate using models from the Coupled Model Intercomparison Project phase 5 (CMIP5) (Taylor et al. (2012)), while Mamalakis et al. (2021) found that the inter-model spread persists in CMIP6 models (Eyring et al. (2016), O'Neill et al. (2016)).

Previous research has established that the meridional position of the ITCZ in the mean climate is influenced by the inter-hemispheric energy transport (Donohoe et al. (2013), Donohoe (2016), Donohoe and Voigt (2017)). Given its influence on cross-equatorial heat transport, the Atlantic Meridional Overturning Circulation (AMOC) is thus believed to play a pivotal role in governing the ITCZ's position and meridional displacements (Frierson et al. (2013), Marshall et al. (2014), Schneider et al. (2014), Yu et al. (2019), E. Moreno-Chamarro et al. (2020)). In fact, in the mean climate the AMOC transports ~0.5 PW of heat into the Northern Hemisphere (Buckley and Marshall, 2016). In addition, several prior studies have employed climate models to conduct ad-hoc experiments to investigate the impacts of reduced energy transport by the AMOC on precipitation. These studies demonstrated that a weakened AMOC results in reduced precipitation over the North Atlantic and a southward shift of the ITCZ's zonal mean position over the tropical ocean (e.g., Vellinga and Wood (2002), Zhang and Delworth (2005), Stouffer et al. (2006), Parsons et al. (2014), Jackson et al. (2015), Good et al. (2021), Orihuela-Pinto et al. (2022)). A southward shift of the ITCZ balances the reduced northward heat transport by the AMOC, and is associated with a meridional displacement of the Hadley Cells and an intensification of the ascending branch of the Northern Hemisphere Hadley Cell (Zhang and Delworth 2005, Bellomo et al. 2023, Lionello et al., 2024). Consequently, the response of the AMOC to climate change might significantly impact the tropical Atlantic ITCZ in future climate scenarios, considering the general consensus among climate models on a substantial weakening of the AMOC by the 21st century (Weijer et al. (2020), Rahmstorf et al. (2015), Caesar et al. (2018), Bellomo et al. (2021)). In particular, Bellomo et al. (2021) investigated abrupt 4x$CO_2$ simulations from the CMIP5 and CMIP6 archives and demonstrated that a sufficiently strong decline in the AMOC intensity can modulate the global precipitation response. Furthermore, idealized experiments with one global climate model showed that a weakened AMOC, relative to rising greenhouse gases, can lead to a southward shift of the tropical Atlantic ITCZ in the annual mean (Liu et al. 2020, Bellomo and Mehling 2024).

The specific role that the representation of the AMOC plays in the large inter-model spread in Atlantic ITCZ projections of the current century has remained largely unexplored, despite the potential significance of the AMOC's response to increasing greenhouse gases (Wang et al. (2014)). In this study we address this gap by employing a set of 30 latest-generation global climate models to examine the ssp5-8.5 high-emission scenario, where the AMOC undergoes substantial weakening by 2100, though the extent of its weakening is uncertain across models. Through this multi-model approach, we group models based on the magnitude of AMOC decline, revealing that the spread in the representation of AMOC decline is tightly linked to inter-model differences in anticipated changes of both tropical rainfall and mean atmospheric circulation over the Atlantic sector. Nevertheless, in these simulations it is challenging to isolate the precipitation response due to AMOC decline from that due to other climate feedbacks related to anthropogenic forcing and ongoing climate change. Therefore, to build consistency on our findings, we supplement our analysis with idealized model experiments using EC-Earth3, a global climate model that participates in CMIP6.

## 2. Materials and Methods

### 2.1 Model simulations

We examine the annual mean tropical Atlantic rainfall response to the AMOC decline in an ensemble of 30 model simulations from the CMIP6 archive (Eyring et al. (2016)). We compare the shared socio-economic pathway ssp5-8.5 experiment with the historical run. The historical experiment was forced with observed estimates of human-induced and natural radiative forcings from 1850 to 2014 (O' Neill et al. (2016)). The ssp5-8.5 future projection spans the years between 2015 and 2100, and was forced with estimates of greenhouse gas concentrations assuming a high emission scenario reaching a radiative forcing at the top of atmosphere of 8.5 W/m$^2$ by the end of the 21$^{st}$ century. The ssp5-8.5 experiment was initialized from the last year of the historical experiment. We analyze all 30 models that made available the variables needed for this study. Where available, we pick the r1i1p1f1 ensemble member. The list of models and the relative ensemble member is provided in Table 1.

In addition to the ssp5-8.5 simulations, we analyze two experiments carried out with the global climate model EC-Earth3. These experiments were started from the preindustrial control ("piControl") experiment (ensemble member r1i1p1f1) from the CMIP6 archive. The piControl experiment spans 500 years during which all the external radiative forcings are maintained at pre-industrial levels (CO2 at ~ 284 ppm). In the first experiment, the CO2 was abruptly quadrupled from the mean annual preindustrial level and was held fixed for 150 years ("abrupt 4x$CO_2$") (Eyring et al., 2016). This experiment is stored in the CMIP6 archive (ensemble member r81i1p1f1). The second one is an ad-hoc experiment that aims at maintaining the AMOC strength fixed at the preindustrial level, despite an abrupt fourfold increase in atmospheric CO2 applied as external forcing. This was achieved by adding a virtual salinity flux poleward of 50°N in the North Atlantic and Arctic Oceans. We ran a small ensemble of 3 members for the second experiment. Although the 3 members share the same configuration as the EC-Earth3 version used in CMIP6 and initial conditions, they were forced with salinity fluxes of slightly different magnitudes (+0.4Sv, +0.5Sv, and +0.6Sv) to span the model's internal variability of AMOC strength in the preindustrial simulation (Meccia et al. 2022). We analyzed each member individually and found no large differences among them. Therefore, only results from the ensemble mean of the 3 members are presented hereafter, referred to as the "PI-fixed AMOC" experiment. The abrupt4x$CO_2$ and PI-fixed AMOC experiments allow us to separate the response of tropical rainfall to $CO_2$ forcing, from the response to an AMOC weakening due to the same forcing, thereby isolating the impacts of the AMOC weakening on tropical rainfall. Note that these experiments are identical to those shown in Bellomo and Mehling (2024) and we refer the reader to that publication for further technical details. Timeseries of AMOC strength in the ssp5-8.5 and EC-Earth3 idealized experiments are shown in fig. 1.

### 2.2 Data analysis and statistical methods

For the 30 ssp5-8.5 simulations, we define the control climatology as the time-mean from the whole historical simulations (165 years spanning from 1850 to 2014). For each model, we compute changes as the difference between the mean of years 2071 through 2100 (30 years) of the ssp5-8.5 simulation and the control climatology. To ensure changes do not lie within the natural climate variability, we perform a Student's t-test between the historical and the future timeseries, using a significance level of 90%. All model output is interpolated to a common 2.5° x 2.5° latitude/longitude grid before performing the analysis.

To assess the impacts of the AMOC decline in the CMIP6 ssp5-8.5 scenario, relative to other changes, we create 2 groups of 10 models each (out of the 30 models employed in this study), based on the amount of the AMOC decline with respect to the control climatology (see Tab.1). The models belonging to the group in which the AMOC declines less are hereafter referred to as "SAD" (small AMOC decline), while the models belonging to the group in which the AMOC declines more are hereafter referred to as "LAD" (large AMOC decline). The remaining 10 models display intermediate values of AMOC decline, and do not belong to any group. In this study, we define the AMOC strength index as the maximum value of the overturning streamfunction at 26.5 °N and below 500m depth (Buckley and Marshall 2016). Fig. 1 shows timeseries of AMOC strength, and models are coded with line types based on which group they belong to. Detailed information on AMOC strength changes in each model is also provided in Table 1. The division into SAD and LAD groups does not change if the reference latitude used to compute AMOC index is varied in the range 20°N to 50°N. To test inter-model agreement and assess whether differences in simulated climate change are statistically significant between "LAD" and "SAD" groups, we perform a Student's two-tailed t-test in the spatial maps. Where the test indicates that the differences are statistically significant, we argue that the driver of the differences between the two groups is the different amount of the AMOC decline. Further, we normalize changes in precipitation and atmospheric circulation of each model by its corresponding global mean near-surface temperature change (ΔGMTAS), so that changes are shown per degree of global warming. We note that dividing by ΔGMTAS also reduces the influence of the global climate feedback on our results, helping to isolate the effect of AMOC on ITCZ changes by mitigating the response to increased $CO_2$. Thus, where statistically significant, we deem the differences in simulated precipitation impacts to arise from the difference in AMOC decline (c.f. Bellomo et al., 2021). Alternatively, we also conduct a bootstrap test to further check whether the differences between the two groups of models are significantly different from the 1000 differences of two randomly chosen groups, and results are consistent with the t-test.
For what concerns the abrupt $4xCO_2$ and the PI-fixed AMOC experiments with EC-Earth3, we compute changes as the difference between time means over model years 50 to 150, covering years in which the weakened AMOC strength is in nearly steady, and the corresponding control climate, here defined as long-term means of the whole piControl run (500-year period). Again, we perform a Student's t-test (at 90% significance level) between the control and the idealized experiments timeseries, to ensure that changes are not part of the model's internal climate variability. Time series of AMOC strength in the EC-Earth3 experiments are also shown in fig. 1. While in the ssp5-8.5 simulations, the AMOC decline was a consequence of CO2 increase, in these idealized experiments we attribute the precipitation impacts to the AMOC weakening by computing the difference between the $4xCO_2$ and the PI-fixed AMOC experiment. To be consistent with the analysis of the ssp5-8.5 simulations, we also normalize changes in ad-hoc EC-Earth3 experiments by ΔGMTAS. We perform a Student's t-test between the abrupt $4xCO_2$ and the PI-fixed AMOC model's years (100 years period, from years 50 to 150) to ensure differences in simulated climate between the experiments do not derive from the EC-Earth3 model's internal climate variability.

To investigate the role of a weakening AMOC in shifting the ITCZ in the meridional direction in the ssp5-8.5 simulations, we evaluate the correlation coefficient and the strength of the linear regression between the magnitude of the AMOC decline and the difference in net precipitation (P – E) anomalies in two adjacent areas in the Atlantic deep tropics, where the P – E change differences between the models are the strongest (see boxes in fig. 3d). These regions are meridionally displaced and have coordinates (30°W – 8°W, 12°N – 3°N) and (47°W – 25°W, 3°N – 6°S). We compute the differences as the area-averaged P – E anomalies in the southern box minus those in the northern box. Thus, positive values indicate a southward migration of the climatologically wetter region, while negative values indicate an intensification of the maximum of P – E. Our results are similar if we choose slightly different regions for the two boxes.

| Model | Variant label | ΔAMOC [Sv] | ΔAMOC [%] | AMOC mean strength [Sv] | ΔGMTAS [°C] | Assigned group |
|---|---|---|---|---|---|---|
| FGOALS-f3-L | r1i1p1f1 | -4.34 | -21.94 | 19.78 | 4.09 | SAD |
| MPI-ESM1-2-LR | r1i1p1f1 | -4.651 | -24.73 | 18.80 | 3.46 | SAD |
| INM-CM5-0 | r1i1p1f1 | -5.11 | -25.97 | 19.69 | 3.17 | SAD |
| ACCESS-ESM1-5 | r1i1p1f1 | -5.21 | -28.79 | 18.09 | 4.26 | SAD |
| CanESM5-CanOE | r1i1p2f1 | -5.45 | -44.44 | 12.26 | 6.37 | SAD |
| MIROC-ES2L | r1i1p1f2 | -5.563 | -44.52 | 12.49 | 3.59 | SAD |
| MPI-ESM1-2-HR | r1i1p1f1 | -5.458 | -32.15 | 16.97 | 3.38 | SAD |
| CanESM5 | r1i1p1f1 | -5.67 | -45.56 | 12.46 | 6.25 | SAD |
| CAS-ESM2-0 | r1i1p1f1 | -5.82 | -30.11 | 19.33 | 4.45 | SAD |
| MIROC6 | r1i1p1f1 | -6.07 | -42.94 | 14.14 | 3.36 | SAD |
| UKESM1-0-LL | r1i1p1f2 | -6.184 | -40.10 | 15.42 | 6.06 | No group |
| INM-CM4-8 | r1i1p1f1 | -6.197 | -31.22 | 19.84 | 3.32 | No group |
| CMCC-CM2-SR5 | r1i1p1f1 | -6.36 | -36.47 | 17.44 | 4.70 | No group |
| HadGEM3-GC31-LL | r1i1p1f3 | -6.545 | -39.79 | 16.44 | 5.70 | No group |
| CMCC-ESM2 | r1i1p1f1 | -6.73 | -38.3 | 17.57 | 4.63 | No group |
| CNRM-CM6-1-HR | r1i1p1f2 | -6.86 | -49.23 | 13.93 | 5.05 | No group |
| ACCESS-CM2 | r1i1p1f1 | -7.23 | -39.25 | 18.43 | 4.97 | No group |
| CNRM-ESM2-1 | r1i1p1f2 | -7.54 | -45.03 | 16.74 | 4.49 | No group |
| CNRM-CM6-1 | r1i1p1f2 | -7.87 | -48.13 | 16.35 | 5.01 | No group |
| GFDL-CM4 | r1i1p1f1 | -8.40 | -45.85 | 18.32 | 4.28 | No group |
| HadGEM3-GC31-MM | r1i1p1f3 | -8.534 | -49.75 | 17.15 | 5.53 | LAD |
| GFDL-ESM4 | r1i1p1f1 | -8.55 | -45.96 | 18.61 | 3.30 | LAD |
| CESM2-WACCM | r1i1p1f1 | -9.80 | -53.21 | 18.42 | 5.10 | LAD |
| CESM2 | r10i1p1f1 | -10.32 | -55.15 | 18.72 | 4.94 | LAD |
| FGOALS-g3 | r1i1p1f1 | -10.60 | -37.75 | 28.07 | 3.27 | LAD |
| GISS-E2-2-G | r1i1p3f1 | -10.74 | -43.71 | 24.57 | 3.69 | LAD |
| MRI-ESM2-0 | r1i1p1f1 | -11.587 | -62.82 | 18.44 | 3.94 | LAD |
| NorESM2-MM | r1i1p1f1 | -12.12 | -56.34 | 21.51 | 3.43 | LAD |
| NorESM2-LM | r1i1p1f1 | -12.15 | -57.98 | 20.96 | 3.33 | LAD |
| GISS-E2-1-G | r1i1p1f2 | -13.44 | -55.24 | 24.33 | 4.21 | LAD |

**Table 1:** *List of models: model's name, variant label (ensemble member), AMOC change (in Sv and percentage) between the ssp5-8.5 simulations and the control climatology, AMOC historical mean strength, ΔGMTAS, belonging group.*

**2.3 Moisture budget**

To investigate the mechanisms driving changes in P – E we perform the atmospheric moisture budget as derived in D'Agostino and Lionello (2020). Here we provide only a short overview of this method, pointing the reader to D'Agostino and Lionello (2020) and references therein for a more in-depth derivation. To compute the moisture budget, the anomalies are computed by monthly mean outputs, the following equation is used:

$$\rho_w g \delta(\bar{P} - \bar{E}) \approx \underbrace{-\int_0^{\overline{p_s}} (\overline{\boldsymbol{u}}_{control} \cdot \nabla \delta \bar{q} + \delta \bar{q} \nabla \cdot \overline{\boldsymbol{u}}_{control}}_{\Delta TH} +$$

$$\underbrace{-\int_0^{\overline{p_s}} (\delta \overline{\boldsymbol{u}} \cdot \nabla \overline{q_{control}} + \overline{q_{control}} \nabla \cdot \delta \overline{\boldsymbol{u}}}_{\Delta DY} + (\delta TE + \delta NL) + \delta S \quad (1)$$

Where P is precipitation, E is evaporation, **u** is the wind vector, q is specific humidity, $p_s$ is the surface pressure field, $\rho_w$ is the density of water, and g is the gravity acceleration. Overbars indicate monthly means and primes indicate deviations from the monthly mean. Here, $\delta$ denotes differences between the simulations (i.e., ssp5-8.5, PI-fixed AMOC, abrupt $4xCO_2$) and their reference control climatology (i.e., historical, piControl).

Through equation (1) we can decompose P - E change as the sum of four drivers. The thermodynamic contribution (ΔTH) reflects changes in the mean humidity and includes changes of humidity gradient along the direction of the mean flow and changes of mean humidity in areas of mean flow convergence or divergence (i.e, constrained by the Clausius-Clapeyron relation). The dynamic contribution ΔDY involves changes in the mean meridional circulation and moisture transported by the mean wind flow. The transient eddies (ΔTE) and nonlinearities (ΔNL) are instead due to changes in the covariance of sub-monthly humidity and wind anomalies. Since the ΔTE + ΔNL term cannot be computed explicitly from monthly data we have used the approach of D'Agostino and Lionello (2020), where it is estimated as:

$$\Delta TE + \Delta NL \ = \rho_w g \delta(\bar{P} - \bar{E}) - \ \delta\left[\nabla \int_0^{\overline{p_s}} \bar{\boldsymbol{u}} \cdot \bar{q} \ dp\right] \ (2)$$

while the surface contribution (ΔS) is computed as the residual:

$$\Delta S \approx \ -\delta\left[\nabla \int_0^{\overline{p_s}} \bar{\boldsymbol{u}} \cdot \bar{q} \ dp\right] - \delta TH - \ \delta DY \ (3).$$

To estimate the AMOC influence on each of these contributions to the total P - E change, we compute the anomalous moisture budget for both the SAD and LAD groups and the PI-fixed AMOC and the $4xCO_2$ experiments; then, we analyze the respective differences. Concerning the SAD and LAD groups, we consider changes significant if at least two-thirds of the models agree in the sign of change. We note that for the GISS-E2-1-G we encountered unrealistic fluxes over land, hence for this model we included only values over the ocean.

## 2.4 Tropical Atlantic Meridional Atmospheric Overturning Circulation

We investigate the dynamic contribution of the atmospheric mean flow over the Tropical Atlantic Ocean with an analysis of the atmospheric mass streamfunction, which is closely related to the mean Hadley circulation. To compute the meridional streamfunction in a regional domain (i.e. in the Atlantic sector), the assumption of nondivergent meridional circulation, which is essential to ensure mass conservation in the computation of the Stokes streamfunction, is invalid. Thus, we follow the approach of Zang and Wang (2013), Schwendike et al. (2014) and Nguyen et al. (2018), to partition the three-dimensional circulation into two divergent circulations lying in orthogonal planes that satisfy continuity independently. Through the Helmholtz decomposition, we extract the irrotational component of the meridional wind flow (i.e., the only contribution to the vertical flow), which can be regarded as the meridional overturning circulation. The Hadley overturning can be represented by a streamfunction computed from the divergent meridional wind similar to the Stokes streamfunction:

$$\psi_y(y,p) = \frac{2\pi R \cos y}{g} \int_{p_t}^{p_s} [v_d(y,p)]\, dp$$

where R is Earth's radius, g is the standard gravity, $p_t$ is pressure at the top of troposphere, $p_s$ is the surface pressure field, $v_d$ is the divergent meridional wind, and $y$ is latitude. Brackets refer to a zonal average over a specified limited domain. In this study we select longitudes ranging between 60 °W and 10 °E.

## 3. Results

### 3.1 AMOC circulation change

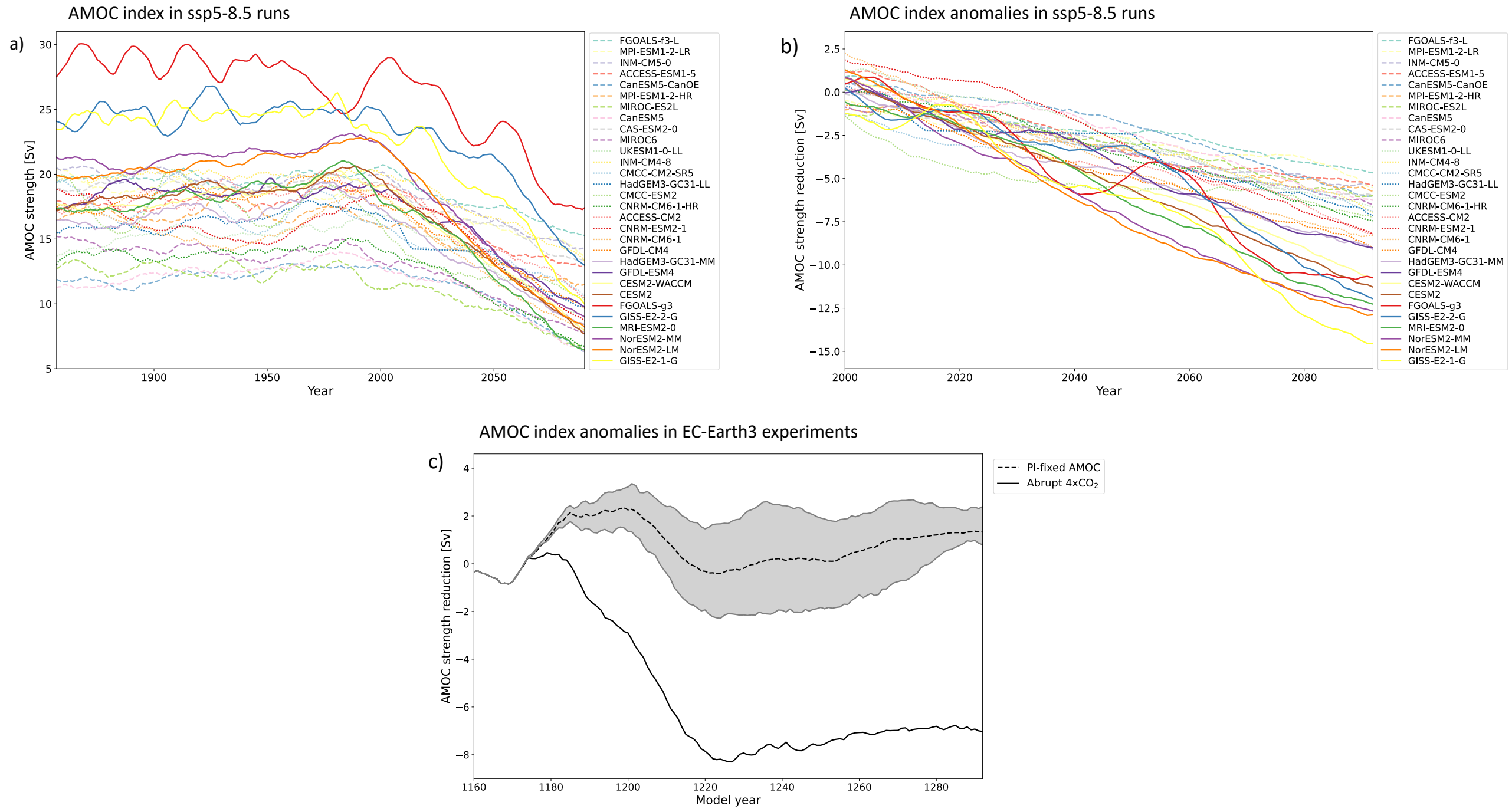

*__Fig. 1 AMOC strength time series at 26.5 °N.__ Panel a shows the annual mean AMOC strength index between years 1850 and 2090 in the ssp5-8.5 simulations, while panel b shows*

*the corresponding annual mean AMOC strength index change with respect to the climatology. Solid curves represent models in the LAD group. Dashed curves represent models in the SAD group. Dotted lines represent models not belonging to any group. In panel c, black curves represent the AMOC index anomalies in EC-Earth3 abrupt 4x$CO_2$ (solid) and PI-fixed AMOC (dashed) experiments. The mean AMOC strength in the picontrol run is 17.8 Sv ±2.4 Sv. The gray shading around the black dashed curve represents the PI-fixed AMOC ensemble spread. A 15-year running mean has been applied to all time series. The AMOC index has been calculated as the maximum value of the overturning streamfunction at 26.5°N and below 500m of depth.*

The timeseries of the annual mean AMOC index at 26.5°N in all the ssp5-8.5 simulations reveal a substantial AMOC weakening occurring by the end of the current century, although a great inter-models spread exists (fig. 1). The highest AMOC decline corresponds to a change in strength of about -63%, and the lowest change is about -22%. The mean change of all models corresponds to about -42% of AMOC strength reduction, with a median of -44%. In Figure 1, solid curves represent the models in which the amount of AMOC reduction belongs to the lower tercile of the distribution of all models (LAD models, AMOC reduction ranges from -8.5 Sv and -13.4 Sv). The dashed curves, on the other hand, represent the models that belong to the upper tercile (SAD models, AMOC reduction ranges from -4.3 Sv and -6.1 Sv). Dotted colored lines represent models in the middle of the range of AMOC declines and do not belong to any groups. Note that models belonging to the LAD group (i.e. models in which the AMOC declines more) also exhibit a stronger AMOC in the climatology (c.f. solid contours in fig. 2), which is consistent with previous studies (e.g., Weijier et al. 2020). As expected, the EC-Earth3 PI-fixed AMOC experiment (dashed black curve) is the only simulation where the AMOC strength, despite increasing greenhouse gases in the background, does not decrease. On the contrary, in the EC-Earth3 abrupt 4x$CO_2$ simulation (solid black curve) the AMOC decreases significantly (~ -7 Sv of change, corresponding to a 43% reduction). A decline of 7 Sv is comparable with that of models belonging to no group (see Table 1).

Figure 2 shows the changes in the meridional oceanic mass streamfunction in a transect of the Atlantic ocean between 30 °S and 70 °N in the SAD group (panel a), in the LAD group (panel b), in the PI-fixed AMOC experiment (panel c), and in the abrupt 4x$CO_2$ experiment (panel d). The reduction of the AMOC circulation strength in the LAD group is much larger than in the SAD group everywhere, and statistically significant at 90% level. The transect shown in panel c shows that the AMOC circulation does not undergo any notable reduction in the PI-fixed AMOC experiment, differently from the abrupt 4x$CO_2$ experiment, where the magnitude of the AMOC reduction is strong and extends to the entire Atlantic basin.

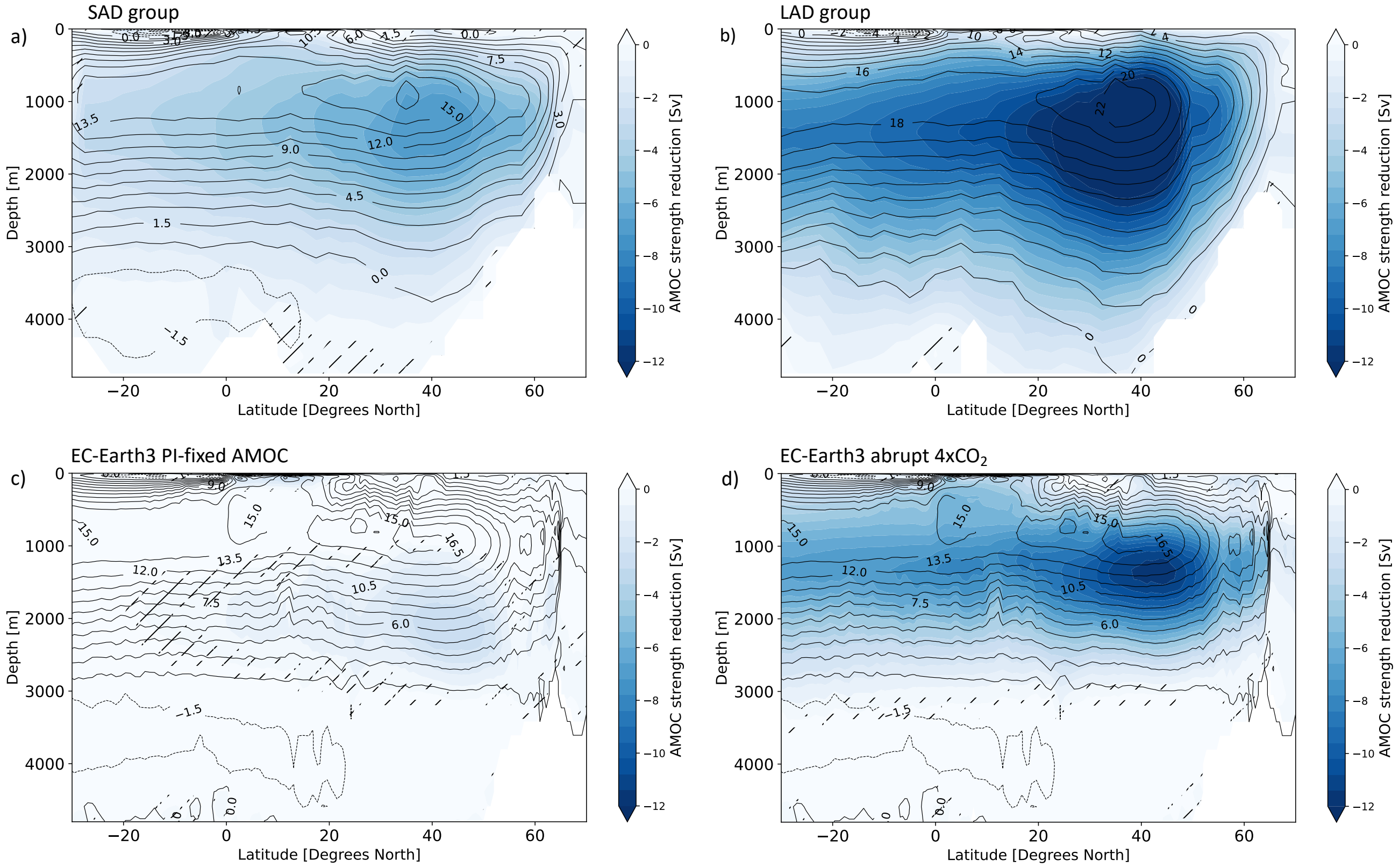

**Fig. 2 Oceanic mass streamfunction change.** *The four panels show changes in the oceanic mass streamfunction in the Atlantic sector. In panels a and b changes are evaluated between the ssp5-8.5 simulations and the historical simulation. Black contours represent the respective climatologies for the two groups of models. Panels c and d show the AMOC circulation change between the PI-fixed AMOC and abrupt 4x$CO_2$ experiments and the piControl run. Black contours represent the control climatology. In all panels, hatches indicate where the statistical t-test on the ocean mass streamfunction change with respect to the reference climatology is not statistically significant at 90% confidence level.*

### 3.2 Net precipitation change in SAD and LAD groups

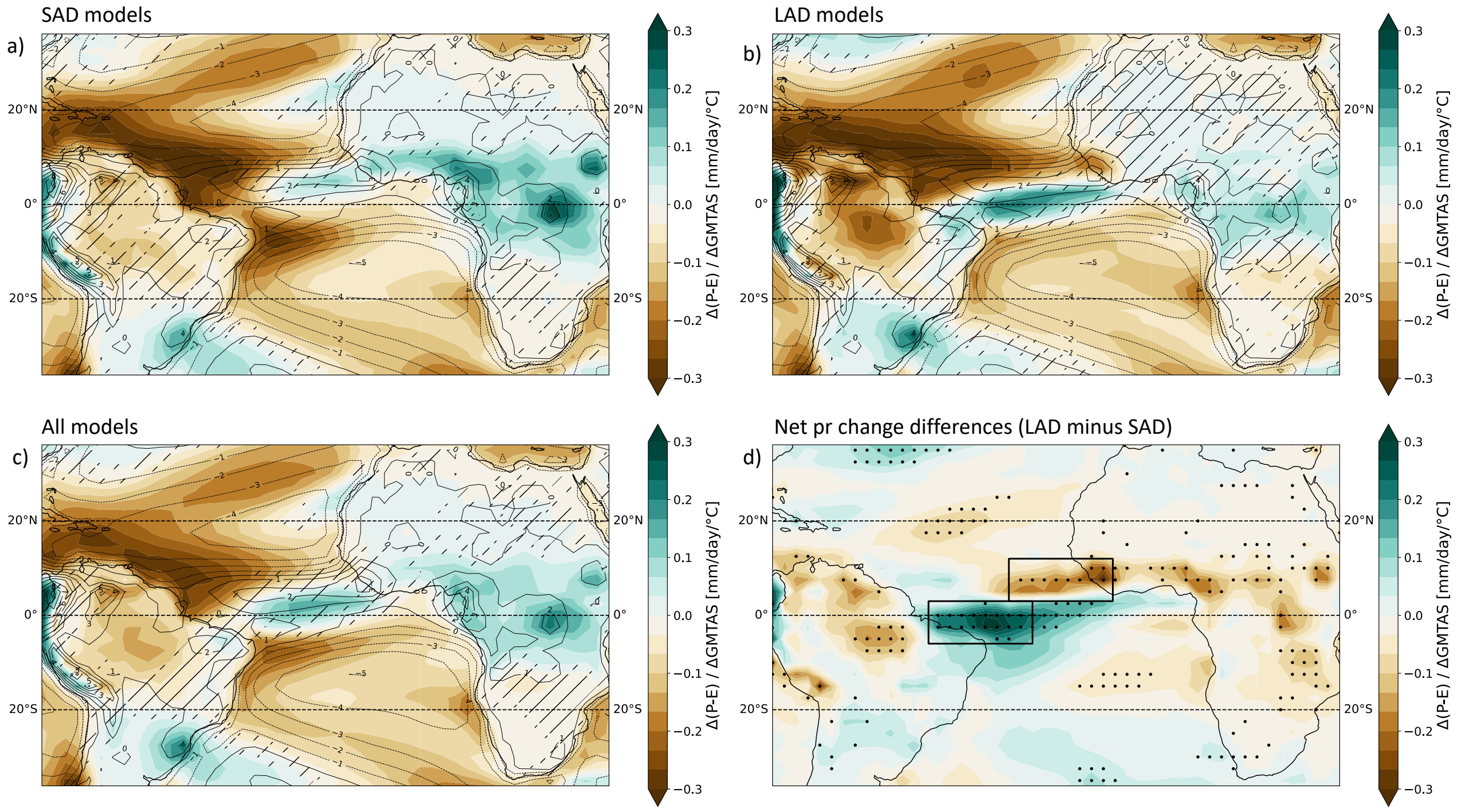

***Fig. 3 Normalized annual-mean net precipitation (P - E) change in ssp5-8.5 scenario projection.*** *Panels a, b and c display the normalized annual mean P - E change for the SAD group, the LAD group, and all models, respectively. Superimposed contours show the mean annual P – E climatology computed as the ensemble mean of the historical run of all models. Hatches indicate where the statistical t-test between the historical and future P - E change is not statistically significant with 90% confidence level of a Student's t-test. Panel d shows the difference between the LAD and the SAD P – E change (Panel b minus Panel a). In panel d stippling indicates where P – E change differences between the LAD and SAD ensemble means are statistically significant with 95% confidence level of a Student's t-test.*

In the SAD group (Figure 3a) P – E intensifies over the ocean close to the western African coast, but decreases all over the Atlantic sector and Amazon. Overall, models with a small AMOC decline project a drying tendency of the climatological ITCZ (here intended as the tropical region where the annual-mean moisture convergence occurs, i.e. the moisture ITCZ (Schneider and Byrne (2016)), over the western side of the Atlantic basin and an intensification over the eastern side at positive latitudes resulting in a drier tropical Atlantic equator. We note that this thermodynamic constraint is less robust over land, as reported in previous studies (Byrne and O' Gorman (2015)). In contrast, the LAD group (Figure 3b) exhibits a southwards shift in P – E, featuring increased P – E right at the equator and drying tendencies at higher latitudes. The maximum of the tropical Atlantic precipitation change (Figure 3b) is displaced westwards with respect to the climatological maximum, consistent with previous studies suggesting it could be related to the Eastern Pacific Ocean anomalous warming under future climate change (Nicknish et al. (2023)). Hence, when the multi-model mean is taken (Figure 3c) the change in net precipitation shows a mix of these two patterns. P – E shows a slight increase below the climatological maximum, specifically between the equator and 4°N, while it decreases in the subtropics. Figure 3d displays the LAD minus SAD P – E change, which better highlights the role of a relatively larger AMOC decline. LAD models show a substantially drier tropical ocean between 4 °N and 10 °N. At the same time, a large band of positive values from 4 °N towards southern latitudes indicates a diffused wetter climate for the LAD group relative to the SAD group. Major differences occur in the deep tropics, further confirming a southward ITCZ migration in the LAD group, which seems to be absent in the SAD group.

Although the model grouping is based on the reduction of mass transport (i.e. strength of the overturning mass streamfunction change at 26.5°), further analyses (not shown) suggest that both model groups experience a reduction in oceanic heat transport, though this is more pronounced for the LAD group. The lack of an ITCZ shift in SAD models may be explained by other mechanisms constraining its meridional position (Bischoff and Schneider (2014), Schneider et al. (2014)), making the signal unclear when the oceanic heat transport is reduced but only to a limited extent. In contrast, a more significant reduction shows a net southward shift that is consistent across LAD models.

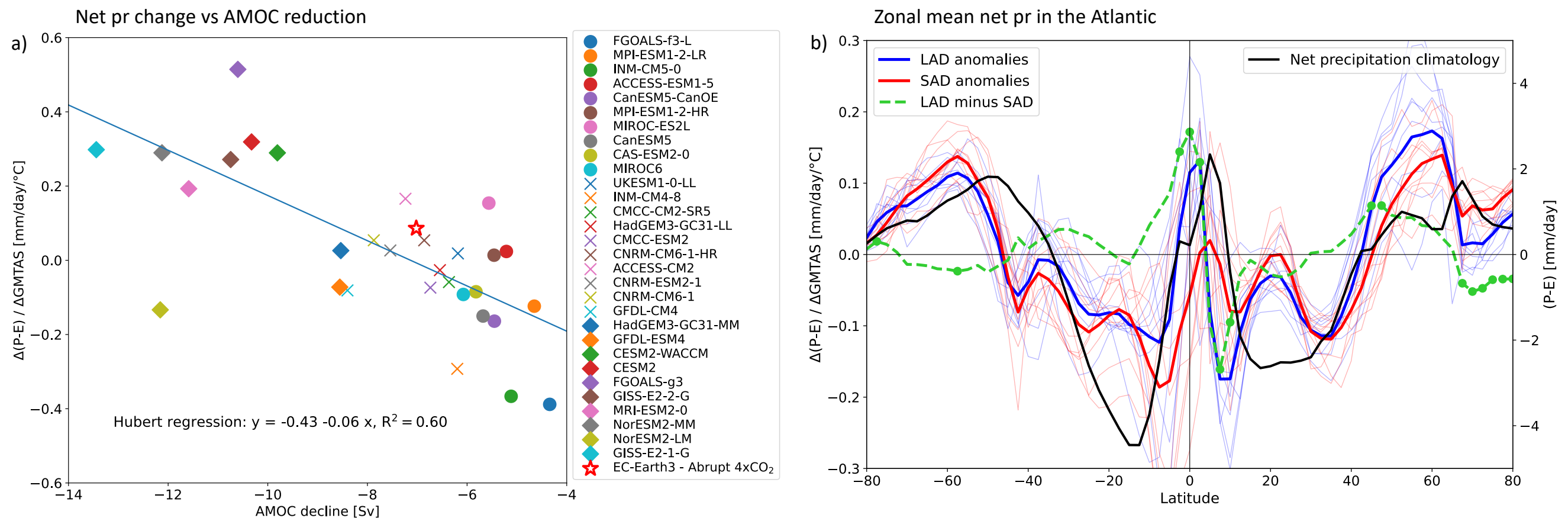

***Fig. 4. Meridional displacement of zonal-mean tropical Atlantic net precipitation in its annual mean in ssp5-8.5 scenario projection.*** *Panel a shows the normalized P - E change difference between the two boxes in Figure 3d (south box minus north box), against the*

*AMOC reduction expressed in Sv. Circles indicate the SAD models, diamonds indicate the LAD models, crosses indicate the other models, while the red star represents the EC-Earth3 abrupt 4x$CO_2$ experiment. The blue line is the Huber regression (that is robust to outliers) on all models excluding the 4x$CO_2$ EC-Earth3 simulation. Panel b displays the zonal mean annual P – E in the Atlantic sector (between 35 °W and 5 °W): the black curve shows the climatological P – E computed from all the historical simulations ; the red curves show P – E change for the members of the SAD group and the SAD group ensemble mean (thicker red curve); the blue curves show P – E change for the members of the LAD group and the LAD group ensemble mean (thicker blue curve); the green curve shows the difference between the two groups. Markers on the green curve indicate where the differences between the SAD and LAD groups are statistically significant with 95% confidence level of a Student's t-test.*

To better investigate the inter-model spread in the ITCZ response, in Figure 4a we show the difference in normalized P – E change between the north and south box (see fig. 3d) against the AMOC change for each model. Positive values in Figure 4a indicate a southward shift of the moisture ITCZ over the Atlantic, while negative values indicate an intensification of the climatology. Most LAD models exhibit positive values, while most SAD models display negative values, indicating a clear division. We use the same boxes to evaluate differences in P – E change in the EC-Earth3 abrupt 4x$CO_2$ experiment (red star in panel 4a) and note that the ITCZ changes fall within the models not belonging to the SAD or LAD groups.

We perform both a robust Hubert regression (y = -0.061 x -0.435 ) with a $R^2$ of 0.60 (Figure 4a, blue line), which excludes outliers, and a linear regression analysis (y = -0.052 x - 0.413, not shown in Figure 4a), with a statistically significant $R^2$ of 0.46, which includes all models except for the abrupt 4x$CO_2$ EC-Earth3 experiment. The regression analysis supports the importance of inter-model differences in AMOC decline in shaping tropical precipitation spatial distribution. Specifically, larger AMOC declines result in a more pronounced southward shift of the moisture ITCZ over the Atlantic, being therefore a fundamental factor in order to understand projected future changes. Moreover, Figure S1 shows P – E change regressed onto the AMOC decline, further supporting this result. We also note that results are similar if we use precipitation instead of P – E (Figure S2).

Figure 4b shows the zonal mean P – E changes for the SAD (thick red curve) and the LAD (thick blue curve) groups, compared to the climatological zonal mean P – E (black curve), averaged over the year and the Atlantic sector. As expected, the climatology displays a peak around 5 °N, corresponding to the annual mean location of the ITCZ. In the SAD ensemble, positive anomalies near the ITCZ suggest a slight strengthening of the climatological peak of P – E in that region. Conversely, a large AMOC decline (LAD) leads to negative anomalies at 5 °N but positive anomalies at the equator, corroborating a southward shift of the Atlantic ITCZ. The zonal mean P – E changes difference (green line, markers indicate statistical significance at each latitude), better shows the substantial role of a stronger AMOC decline in displacing southwards the positive P – E changes in the tropical Atlantic.

### 3.3 Mechanisms of net precipitation change in SAD and LAD groups

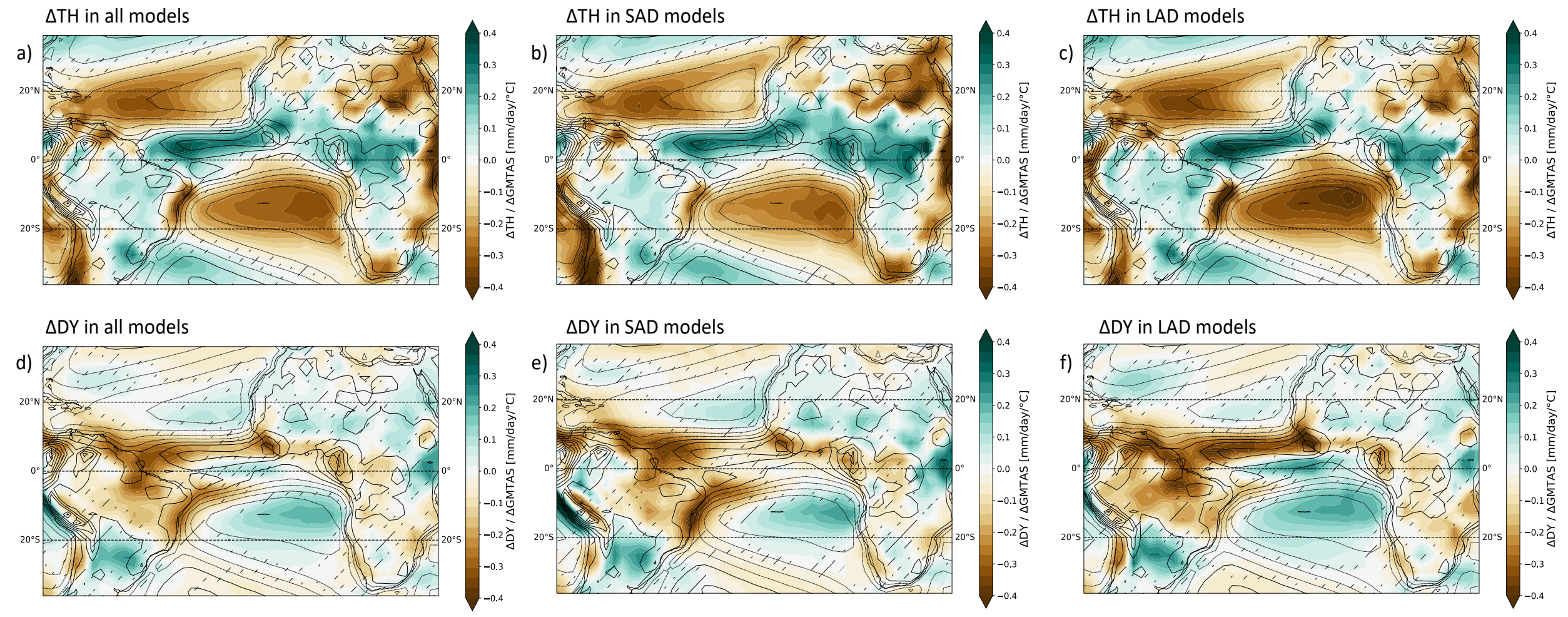

***Fig. 5 Normalized annual-mean thermodynamic and dynamic drivers of net precipitation change in ssp5-8.5 scenario.*** *Panels a, b and c show changes in the thermodynamic contribution (ΔTH) to the moisture budget for all the models, for the SAD group and for the LAD group, respectively. Panels d, e and f show changes in the dynamic contribution (ΔDY) to the moisture budget for all the models, for the SAD group and for the LAD group, respectively. Note that the sum of these two contributions plus the contribution given by transient eddies and surface quantities (provided in SI Figure S3) equals the P – E change displayed in Figure 3 by construction. Hatches indicate regions of inter-model disagreement, that is, where less than two-thirds of the models agree on the sign of the change. Superimposed contours show the mean annual P – E climatology computed as the ensemble mean of the historical run of all models.*

Changes in P – E appear to be driven by different processes depending on how much the AMOC declines. To investigate these mechanisms, we apply an atmospheric moisture budget to model data (c.f. D'Agostino and Lionello 2020). This allows us to separate thermodynamic (ΔTH) from dynamic contributions (ΔDY), and to further quantify changes resulting from transient eddies and nonlinearities (ΔTE + ΔNL). Since ΔTE + ΔNL and the residual are small, in the main text we only show ΔTH and ΔDY, while the other terms can be found in Figure S3. Focusing on changes over the ocean, Figures 5a, 5b and 5c display consistent spatial patterns of P – E change, indicating that the ΔTH term contributes primarily by intensifying the hydrological cycle (e.g., Held and Soden (2006)). The strength of anomalies is slightly different among these panels, although the normalization by ΔGMTAS helps to limit this effect. It is worth noting that LAD models exhibit more warming in the tropical Atlantic compared to SAD models, implicating stronger thermodynamic changes (cf. figs. 5b and 5c). Nevertheless, we can conclude that ΔTH drives precipitation impacts independently of the amount of AMOC weakening in these models.

In contrast, the ΔDY term shows substantial differences among the three panels (figs. 5d, 5e and 5f). In the "all models" panel (Figure 5d), a zonal asymmetry appears between 35°W and 10°W, along with a drying tendency near the coasts of South America. This pattern represents the average effect of various processes. Specifically, the SAD group (Figure 5e) displays a strong dynamical drying signal near the eastern South American coasts, extending throughout the deep tropics of the Atlantic and intensifying towards the Gulf of Guinea. Instead, the dynamic component of the LAD group (Figure 5f) exhibits a meridional dipole-like structure located between the equator and 10°N, resembling the "all models" panel but with stronger anomalies at the equator. Furthermore, we observe similarities between Figures 3a and 5e (P – E changes and DY changes for the SAD group), particularly the drying signal along the American coasts, and between Figures 3b and 5f, showing the (dynamical) equatorial moistening displayed by the LAD ensemble. These findings underscore the substantial impact of dynamic drivers on the P – E changes in the tropical Atlantic. In the case of a smaller AMOC decline, the deep tropical Atlantic experiences dynamic drying, while a larger AMOC decline favors ΔDY-induced moistening of the equator, achieved through the southward migration of the ITCZ (c.f. Figure 4a).

### 3.4 Atlantic atmospheric overturning circulation change in SAD and LAD groups

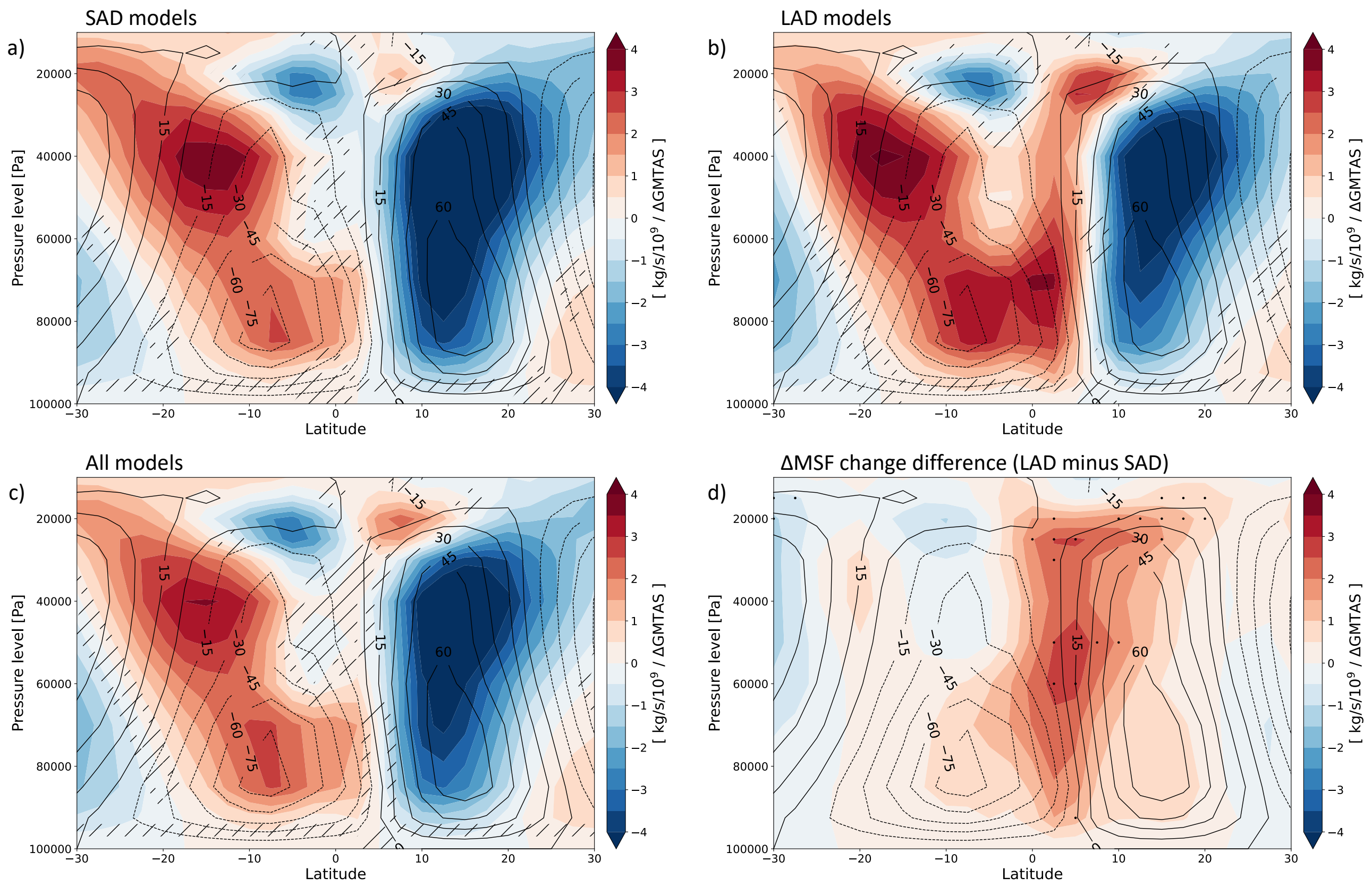

***Fig. 6 Normalized annual-mean Atlantic mass stream function changes in ssp5-8.5 scenario.*** *Panels a, b and c display the normalized annual mean Atlantic atmospheric overturning circulation change for the SAD group, the LAD group, and all models, respectively. The streamfunction is calculated on longitudes ranging from 60 °W to 10 °E. Superimposed contours show climatology computed as the ensemble mean of the historical run of all models. Hatches indicate where the statistical t-test between the control and future P – E distribution is not statistically significant with 90% confidence level of a Student's t-test. Panel d shows the difference between the LAD and the SAD atmospheric circulation anomalies (Panel b minus Panel a). Stippling indicates where atmospheric circulation change differences between the LAD and SAD ensemble means are statistically significant with 90% confidence level of a Student's t-test.*

The dynamic term ΔDY appears to play a prominent role in causing differences in P – E changes between the SAD and LAD groups. Because ΔDY quantifies changes in the column-integrated moisture transported by the mean flow, which in the Tropics mostly occurs in the lower branches of the Hadley cells, here we investigate changes in the meridional overturning atmospheric mass streamfunction across longitudes ranging from 60°W to 10°E. To achieve this, we adopt the approach outlined by Zhang and Wang (2013), and further developed by Schwendike et al. (2015) and Nguyen et al. (2018), which allows us to evaluate the streamfunction over a limited longitudinal range ensuring mass conservation (see Methods). Figures 6a and 6b show the Atlantic mass streamfunction changes (ΔMSF) for the SAD and LAD groups, respectively. Superimposed black contour lines represent the mean atmospheric circulation computed from the historical simulation from all models. The climatological ascending branches of the northern and southern hemisphere annual mean Hadley cells converge at around 5°N, roughly corresponding to the meridional position of mean annual peak of Atlantic precipitation. We now focus on this region and nearby latitudes to explore changes in the mass streamfunction associated with changes in the position of the ITCZ.

In the LAD group (Figure 6b), positive (clockwise) anomalies between the equator and 5°N indicate that the Northern Hemisphere cell strengthens close to the equator whereas the Southern Hemisphere cell weakens. Specifically, the clockwise anomalies at the interface between the two cells suggest a squeezing of the climatologically cross-equatorial mean annual Southern Hemisphere Cell southwards, while the Northern Hemisphere Cell expands. We note that the SAD group (Figure 6a) does not show significant changes in cell intensity in this region. Instead, both the SAD and LAD groups exhibit a broad weakening of the Hadley cells, reflecting the slowdown of the tropical overturning circulation in a warming climate (c.f. Lu et al. 2007, Byrne et al. 2018).

Figure 6d (LAD minus SAD) shows statistically significant differences in the change of the strength of the Atlantic Hadley cells across the entire atmospheric column from the equator to ~ 9°N, extending northward at higher altitudes (i.e. above 30 hPa). Discrepancies at higher altitudes indicate a difference in the rise of the tropopause height. In general, models with stronger AMOC weakening also show more intense tropical warming (Bellomo et al. 2021), resulting in enhanced deepening of tropical convection and heightening of the Hadley cell (Neelin and Held, 1987), hence such significant differences. While differences in the tropopause height changes can be attributed to differences in the tropical warming, we argue that differences in the strength and structure at tropical latitudes and lower elevation of the Atlantic Hadley cells are driven by the inter-model spread in the AMOC response to climate change.

In LAD models, the reduction in cross-equatorial ocean heat transport caused by the weakened AMOC influences the Atlantic Hadley Circulation, possibly leading to a southward shift of the ascending branch in its annual mean (i.e. the Southern Hemisphere Hadley cell tends to become less cross-equatorial). Thus, the dynamical ΔDY contribution to the meridional displacement of P – E change (i.e. the southward shift of the Atlantic ITCZ) may be explained by changes in the Atlantic atmospheric overturning circulation in the group of models featuring a relatively large AMOC decline. We note that when evaluated globally, the atmospheric circulation anomalies show a similar pattern of change, but differences between the groups are less statistically significant (Figure S4), suggesting less influence of the AMOC on the ITCZ on a global scale. Similar changes in the Hadley Cells are consistent with previous studies analyzing idealized experiments where the AMOC is artificially shut down (Orihuela-Pinto et al. (2022), Bellomo et al. (2023)), lending confidence that our results are robust.

### 3.5 Net Precipitation changes in the EC-Earth3 idealized experiments

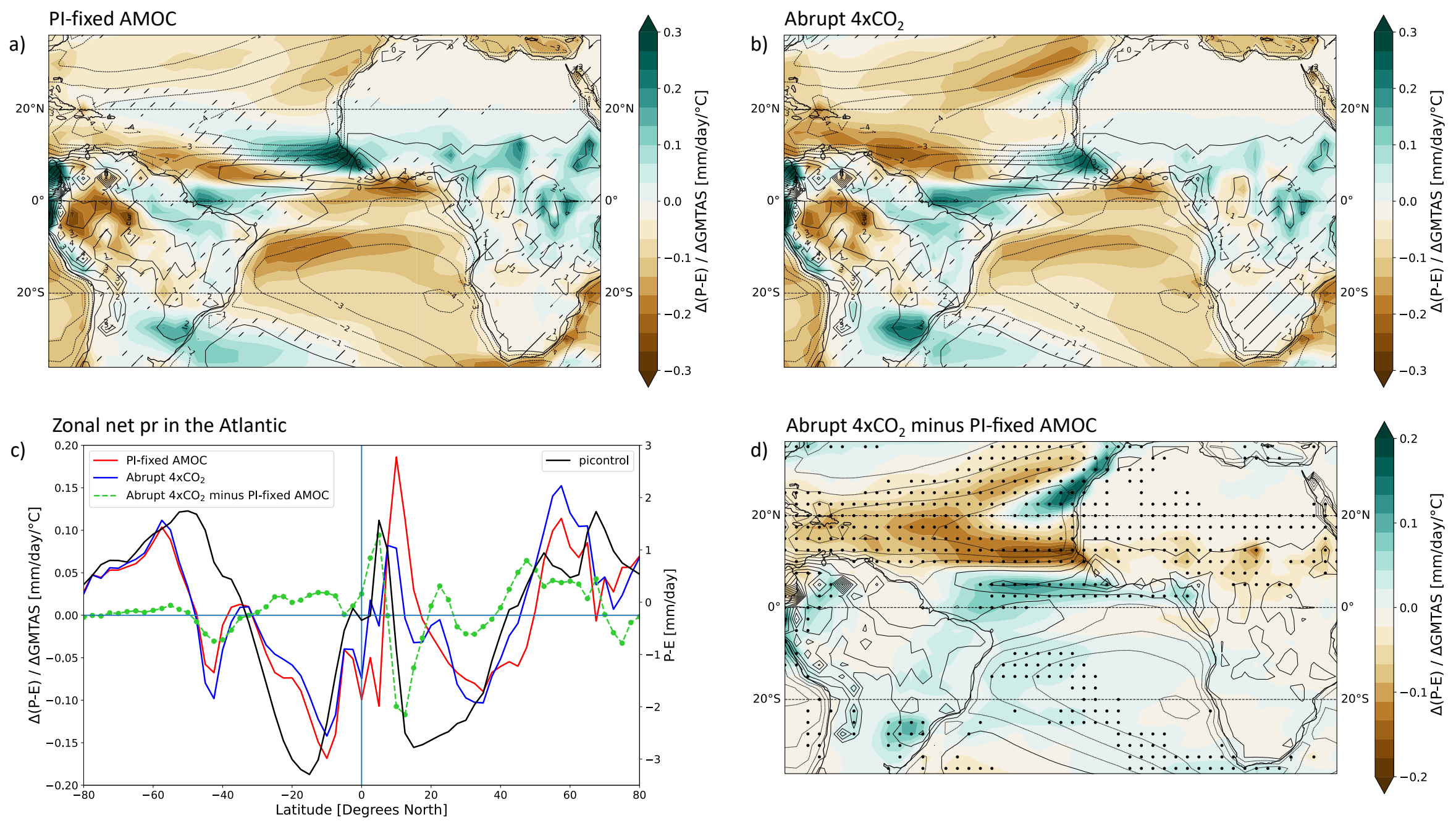

***Fig. 7 Normalized annual-mean net precipitation (P - E) changes in EC-Earth3 idealized experiments.*** *Panels a and b display the normalized annual mean P – E change in the PI-fixed AMOC and abrupt 4x$CO_2$ experiments, respectively. Superimposed contours show the mean annual P – E control climatology. Hatches indicate where the statistical t-test on the P – E change in the experiments is not statistically significant at 90% confidence level. Panel c displays the zonal mean annual P – E in the Atlantic sector (between 35 °W and 5 °W): the black curve shows the piControl zonal mean P – E , the red curve shows the zonal mean P – E change for the PI-fixed AMOC experiment; the blue curve shows the zonal mean P – E change for the abrupt 4x$CO_2$ experiment; the green curve shows zonal mean P – E changes*

*difference between the two experiments (blue curve minus red curve). Markers on the green curve indicate where the differences are statistically significant with 95% confidence level of a Student's t-test. Panel d shows the difference between the PI-fixed AMOC and the abrupt 4x$CO_2$ P – E change (Panel b minus Panel a). Stippling indicates where P – E change differences of the two experiments are statistically significant at 95% confidence level of a Student's t-test.*

Panels 7a and 7b depict P – E changes in the Atlantic sector in the PI-fixed AMOC and the abrupt 4x$CO_2$ experiment, respectively. The changes are normalized by ΔGMTAS, as in Figure 3, to facilitate a comparison between them. P – E changes show a very similar spatial pattern between the two experiments. The moisture ITCZ tends to dry out, while positive anomalies are shifted both northeast and southwest with respect to the climatological peak of P – E under increasing greenhouse gases in the EC-Earth3 model. The zonal mean Atlantic P – E is shifted southwards in the abrupt 4x$CO_2$ run (blue curve in panel 7c) compared to the PI-fixed AMOC experiment (red curve). This is highlighted by the green curve showing the difference between the two. Positive differences between 5°S and 5°N (with a peak between 2.5°N and 5°N) indicate that in the abrupt 4x$CO_2$ the climate is wetter across the equator and southwards with respect to the PI-fixed AMOC run. The green curve clearly resembles the one represented in panel 4b, although anomalies are located at different latitudes. The result corroborates the southwards shift tendency of the anomalous Atlantic ITCZ mainly driven by a substantially weakened AMOC. Since here we analyze one model (EC-Earth3) and in fig. 4b an ensemble of models, some differences are expected. But overall both figures suggest a fingerprint of the AMOC decline on the southward shift of the ITCZ in a warming 21$^{st}$ century. Notably, a major effect of a strong AMOC slowdown as occurs in the abrupt 4x$CO_2$ simulation is the reduced warming in the subpolar North Atlantic (Liu et al. 2020, Bellomo et al. 2021), because the northward transport of heat by the ocean is reduced. Thus, in this experiment, the southward shift of the ITCZ reflects the altered inter-hemispheric energy transport.

Figure 7d (panel 7b minus panel 7a) quantifies the impacts of an AMOC weakening, showing a generally wetter moisture ITCZ and a wetting signal that expands towards the southwest direction (near the South American coast), following the southwest shift already seen and commented in the LAD group anomalies (Figures 3b, 3d). We note that the northeastern wetting seen in panels 7a and 7b is much stronger in the PI-fixed AMOC experiment, and no trace of it remains when assessing the difference between the two, meaning that in the abrupt $4xCO_2$ run it is limited by the declining AMOC. The tongue of positive differences near the north African coast in panel 7d is caused by a strong cooling spot that occurs under the abrupt $4xCO_2$ simulation. The cooling leads to reduced P – E changes because the evaporation diminishes in response to lower temperature (see Bellomo and Mehling (2024))). In fact, it does not appear when considering only precipitation anomalies (see panel d in Figure S5).

### 3.6 Moisture budget in the EC-Earth3 idealized experiments

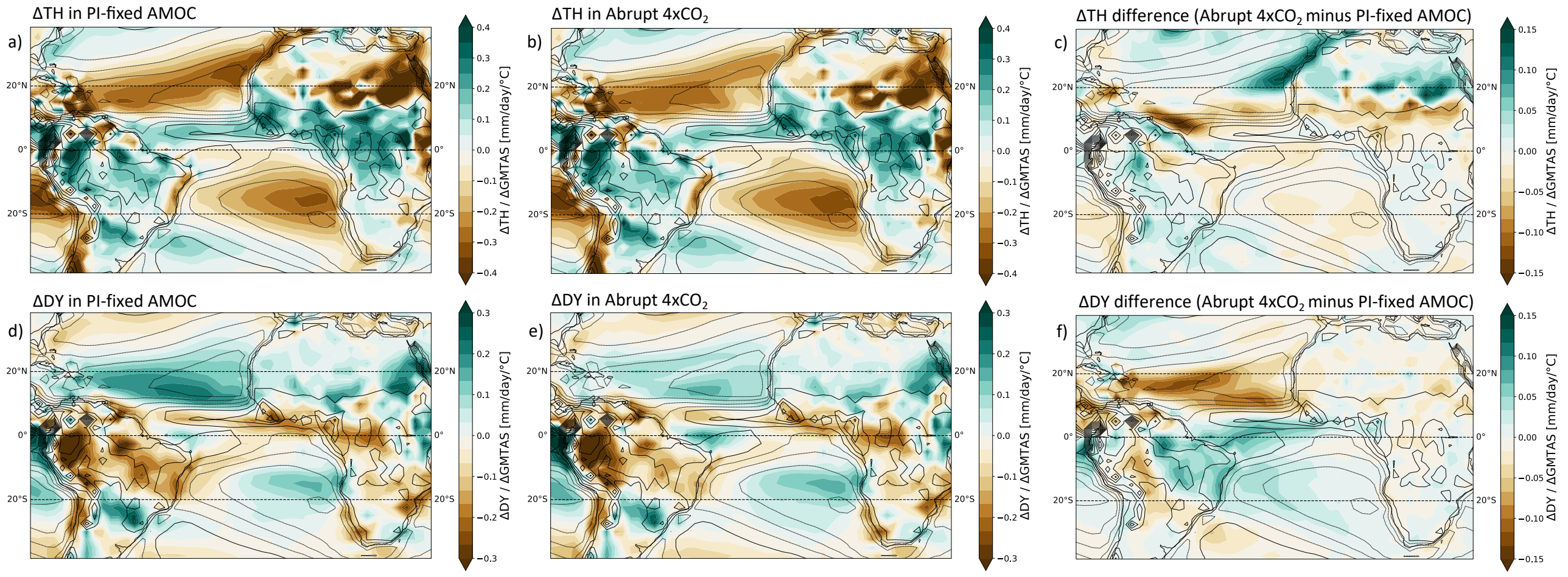

***Fig. 8 Normalized annual-mean thermodynamic and dynamic drivers of net precipitation change in EC-Earth3 idealized experiments.*** *Panels a and b show changes in the thermodynamic contribution (ΔTH) to the moisture budget for the PI-fixed AMOC and the abrupt* $4xCO_2$ *experiment, respectively. Panel c is their difference. Panels d and e show changes in the dynamic contribution (ΔDY) to the moisture budget for the PI-fixed AMOC and the abrupt* $4xCO_2$*, respectively. Panel f is analogous to panel e but for the ΔDY term. Note that the sum of these two contributions plus the contribution given by transient eddies and surface quantities (provided in SI Figure S6) equals the P – E change displayed in Figure 7 by construction. Superimposed contours show the mean annual P – E control climatology.*

As shown before, the ΔTH contribution to the moisture budget is an intensification of the present-day hydrological cycle. It is therefore not surprising that panels 8a and 8b showing the ΔTH contribution in PI-fixed AMOC and the abrupt $4xCO_2$ experiments exhibit a band of positive anomalies slightly north of the equator where the moisture ITCZ over the Atlantic lies in the EC-Earth3 piControl mean climate, and negative anomalies are displaced further north and south of this band. Hence, the ΔTH term (fig. 8c) fails to explain the differences in tropical Atlantic P – E change seen in fig. 7d. Although both the experiments feature the "wet-get-wetter, dry-get-drier" paradigm (Held and Soden 2006), the PI-fixed AMOC experiment reveals a stronger intensification of the hydrological cycle in the Northern Hemisphere. Strong differences are indeed found near the coasts of Northern South America (leading to a tongue of positive differences already mentioned for panel 7c), North Africa and in the African monsoon region. This is caused by the warmer Northern Hemisphere in its annual mean, relative to the Southern Hemisphere, which characterizes the PI-fixed AMOC experiment, as explained earlier. In fact, it is very well-known that a major effect of a strong AMOC slowdown as occurs in the abrupt 4x$CO_2$ simulation is the relative cooling of the Northern Hemisphere (Liu et al. 2020, Bellomo et al. 2021), because the northward transport of heat by the ocean is reduced.

The spatial distribution of the ΔDY contribution to the moisture budget (panels 8d and 8e) is similar in the two experiments. Both panels show a drying signal between 10°S and 10°N that is reminiscent of that seen in the ΔDY contribution derived from the all-models ensemble mean future projections (fig. 5b), except for the positive anomaly being displaced near the South American coast in the EC-Earth3 experiments. Note that in the 4x$CO_2$ this positive signal is broader and more intense between the equator and South America. Fig. 8f shows the ΔDY difference between the abrupt 4x$CO_2$ and PI-fixed AMOC experiments, which allows us to assess the weakened AMOC impact on the column-integrated atmospheric moisture transport. Between the two Hemispheres a bipolar seesaw, with adjacent positive/negative differences that change in sign around 7 degrees north, indicates a stronger dynamical moistening of the equatorial latitudes occurring in the abrupt 4x$CO_2$ experiment. Therefore, fig. 8f demonstrates that an AMOC weakening alone causes changes in the atmospheric circulation that lead to a southward shift of the tropical moisture ITCZ, also in a 4x$CO_2$ forced warmer climate.

## 4. Discussion and Conclusions

Previous studies have found that in climate change projections from the CMIP5 and CMIP6 models, there is large uncertainty in tropical Atlantic precipitation change, including the meridional dislocation of the ITCZ (e.g., Byrne et al. 2018, Mamalakis et al. 2021). One study (Good et al. 2021) based on results from one state-of-the-art climate model (HadGEM3), suggested that inter-model uncertainties in future tropical Atlantic precipitation change may stem from model biases, including the AMOC representation. Other studies have also indicated a potentially significant role for the AMOC in tropical Atlantic precipitation change (e.g., Delworth and Zhang 2005, Jackson et al. 2015, Bellomo et al. 2021, Liu et al. 2020, Nicknish et al. 2023), but our study specifically provides evidence supporting an important role for a weakened AMOC in driving tropical Atlantic P – E changes in a warming 21st century climate. In this study, we used a high-emission scenario (the ssp5-8.5 projections), as the large inter-model spread allowed us to statistically separate the response of the Atlantic ITCZ to different magnitudes of AMOC weakening. However, we would expect similar results even if a less extreme greenhouse gas increase scenario were used, because Weijer et al. (2020) shows that significant inter-model spread exists in the AMOC representation under other emission pathways as well. Because in these simulations the AMOC decline is forced by increasing greenhouse gases, although we took measures to only consider the AMOC impact and not the CO2 forcing, we are not fully able to cleanly separate the effect of greenhouse gases from the effect of the AMOC decline on tropical Atlantic P – E change. To overcome this difficulty, we additionally analyzed ad-hoc experiments with the EC-Earth3 model. In these experiments, we artificially fixed the AMOC strength at pre-industrial levels. With this additional analysis we were able to attribute precipitation impacts to the AMOC decline alone. Our results suggest that the AMOC decline may be a key driver of inter-model spread in meridional displacement of tropical Atlantic ITCZ, and the models featuring a larger AMOC decline in the ssp5-8.5 experiment project a southward ITCZ displacement, while models featuring a smaller AMOC decline exhibit no shift. The EC-Earth3 experiments corroborate findings from the multi-model analysis. The ITCZ over the Pacific basin shows signals of a southward ITCZ shift as well. Nevertheless, the response in the Atlantic is clearer, while the effect in the Pacific is influenced by other effects related to climate change, making it harder to associate the differences in the meridional position to the spread of AMOC decline. Finally, it remains open to explore other competing mechanisms that affect ITCZ displacement under simulation of future climate change, which may also amplify the inter-model spread. Indeed, while AMOC reduction is one factor at play, not all models show a southward shift of the ITCZ, despite all of them exhibiting a decrease in inter-hemispheric ocean heat transport.

To put our results in context, we recall that Byrne et al. (2018) examined an ensemble of 32 model simulations of the 21st century from the CMIP5 archive and found that the median model projected a northward shift of the global mean ITCZ of 0.03° latitude per Kelvin of global-mean surface-air warming, with an interquartile range across models of 0.46°/K. Of the models which they analyzed, 17 out of 32 predicted a northward ITCZ shift by the end of the 21st century, while the remaining models predicted a southward shift. This near-even split between models results in great uncertainty regarding the direction of the ITCZ shift by the end of the 21st century, which has significant implications for regional hydroclimate changes. In our study, we demonstrate that a southward shift of the Atlantic ITCZ is driven by changes in the mean meridional atmospheric circulation, specifically the Atlantic regional Hadley circulation, but only in models featuring a relatively larger AMOC decline. Although we did not specifically examine global-scale ITCZ changes, Figure S4 indicates a southward shift of the Northern Hemisphere Hadley Cell on a global scale in models with relatively larger AMOC declines.

Therefore, when comparing our results with Byrne et al. 2018 and the latest AR6 IPCC report (IPCC 2022), we posit that the inter-model spread in the meridional displacement of the ITCZ may, in part, be attributed to the inter-model spread in AMOC decline. Moreover, beyond meridional shifts, we observe that the AMOC decline also modulates the intensity, spatial structure, and zonal migrations of P – E change. Recent studies (e.g., Nicknish et al., 2023) suggest that AMOC reduction, coupled with eastern equatorial Pacific warming, acts as a key driver of zonal and meridional energy flux shifts, which in turn mirror gross distributional changes of tropical precipitation. Therefore, differences in zonal shifts of P - E between the two groups may arise from inter-model differences in eastern equatorial Pacific warming (c.f. Bellomo et al., 2021), affecting trade wind patterns over the Atlantic. Our findings suggest that, under future global warming, the degree of compensation between thermodynamic (TH) and dynamic (DY) contributions, which is fundamental to understand future African monsoon precipitation (D'Agostino et al. 2019, D'Agostino et al. 2020), is potentially impacted by the AMOC slowdown effect on DY. The ssp5-8.5 simulations also reveal that a significant weakening of the AMOC leads to reduced annual mean rainfall in the Amazon region, which aligns with findings from idealized climate model experiments in which the AMOC is artificially suppressed (e.g., Vellinga and Wood (2002), Parsons et al. (2014), Jackson et al. (2015)). However, the difference in precipitation change simulated by the EC-Earth3 experiments does not agree with this result, therefore caution is needed when attributing it to the declining AMOC.

Finally, we assert that the AMOC plays a crucial role as a key driver of inter-model uncertainty in projections of future tropical Atlantic precipitation changes. Our study shows that targeted experiments are needed to specifically investigate climatic impacts of the AMOC decline relative to anthropogenic global warming. We also suggest that in order to understand the impacts of the AMOC on tropical Atlantic hydroclimate, continued observational campaigns (McCarthy et al., 2015; Frajka-Williams et al., 2019; Li et al., 2021) are essential. An observational constraint on the AMOC response to future climate change would be beneficial for reducing inter-model uncertainty in its impacts.

## Statements and declarations

**Acknowledgements:**

We acknowledge the World Climate Research Programme's Working Group on Coupled Modelling, which is responsible for CMIP, and we thank the climate modeling groups (listed in Table 1 of this paper) for producing and making available their model output.

**Data availability**

CMIP6 data openly available in CMIP6 Earth Science Grid Federation archives at https://esgf-data.dkrz.de/search/cmip6-dkrz/. Relevant dataset generated and analysed for the

Ec-Earth3 experiments during the current study is available at https://doi.org/10.5281/zenodo.11545299

**Funding**

G.C., K.B. and J.v.H acknowledge funding from the European Union Next-GenerationEU within the RETURN Extended Partnership (National Recovery and Resilience Plan – NRRP, Mission 4, Component 2, Investment 1.3 – D.D. 1243 2/8/2022, PE0000005). K.B. acknowledges funding from JPI Oceans and JPI Climate "Next Generation Climate Science in Europe for Oceans"–ROADMAP Project (D. M. 593/2016) and from the European Union's Horizon 2020 research and innovation programme under the Marie Skłodowska-Curie grant agreement No. 101026907 (CliMOC).

**Conflict of interests**

The authors declare that they have no relevant financial or non-financial interests to disclose.

**Author contributions**

K.B., G.C., and J.v.H. conceived the project. G.C. and K.B. retrieved the data. G.C. analyzed the data and prepared the figures. R.D. provided the code to compute the moisture budget and meridional overturning circulation, which was later adapted by G.C. and K.B. for this study. All authors contributed to the discussion and interpretation of the results. G.C. and K.B. wrote the first draft of this manuscript, and all authors contributed to edit and finalize the manuscript. All authors read and approved the final manuscript.

**Code availability**

All code for data analysis associated with the current work can be requested by writing to the corresponding author.